\documentclass[twocolumn,pre,showpacs]{revtex4}

\usepackage{graphicx}
\usepackage{color}
\usepackage{rotating}
\usepackage{amsmath}
\usepackage{amsfonts}
\usepackage{amssymb}
\usepackage{enumerate}
\usepackage{longtable}
\usepackage{dcolumn}
\usepackage{bm}
\usepackage{epsfig}
\usepackage{hyperref}
\begin{document}
\title{Phase change in an opinion-dynamics model with separation of time scales}
\author{Gerardo I\~{n}iguez$^{1}$}
\author{J\'{a}nos Kert\'{e}sz$^{1,2}$}
\author{Kimmo K. Kaski$^1$ }
\author{R. A. Barrio$^{1,3}$}
\affiliation{$^1$BECS, School of Science and Technology, Aalto University, P.O. Box 12200, FI-00076,\email{kimmo.kaski@hut.fi}}
\affiliation{$^2$Institute of Physics and HAS-BME Cond. Mat. Group, BME, Budapest, Budafoki \'{u}t 8., H-1111,\email{kertesz@phy.bme.hu}}
\affiliation{$^3$Instituto de F{\'{\i}}sica, UNAM, Ciudad Universitaria, C.P. 04510, M\'{e}xico D.F.\email{barrio@fisica.unam.mx} }
\date{\textrm{\today}}

\begin{abstract}
We define an opinion formation model of agents in a 1d ring, where the opinion of an agent evolves due to its interactions with close neighbors and due to its either positive or negative attitude toward the overall mood of all the other agents. While the dynamics of the agent's opinion is described with an appropriate differential equation, from time to time pairs of agents are allowed to change their locations to improve the homogeneity of opinion (or comfort feeling) with respect to their short range environment. In this way the time scale of transaction dynamics and that of environment update are well separated and controlled by a single parameter. By varying this parameter we discovered a phase change in the number of undecided individuals. This phenomenon arises from the fact that too frequent location exchanges among agents result in frustration in their opinion formation. Our mean field analysis supports this picture.

\end{abstract}
\maketitle

%\date{Received: date / Revised version: date}

\section{Introduction}
\label{intro}
How are opinions formed? In sociology this is one of the basic questions, but it is also highly relevant for politics, innovation spreading, decision making, and the general well feeling of people \cite{baron2010sps, wu2004sso, watts2007inp}. This complex process depends on various factors or components like confidence, attitudes, communities or media effects \cite{white1976ssm}. Recently, much effort has been invested in modeling different aspects of opinion dynamics and these models are in many ways related to those of physics \cite{castellano2009sps, sobkowicz2009mof}. Unfortunately, the empirical observations are rather sparse. Therefore, the usual strategy is to concentrate on some particular features by making plausible assumptions  for a model, and comparing its results with expectations. Here we will follow this line of study. 

Our starting point is that the comfort feeling of an individual depends on his/her embedding in the society. We get friends mostly with people who are similar to us, share our opinions, tastes etc. In sociology this is called \emph{homophily} and is known to be the major governing principle in friendship formation \cite{baron2010sps}. In terms of physics, this corresponds to ferromagnetic interactions \cite{kadanoff2000sps}. In the language of opinion dynamics this means that: a) The opinion of an individual gets adjusted to that of his/her friendship neighborhood; b) An individual seeks the neighborhood of alike others. Here a) has been the basis of many opinion-dynamics models, both discrete and continuous \cite{holyst2001sim, hegselmann2002odb, holley1975etw, weidlich1991pas, sznajdweron2000oec, deffuant2000mba, vazquez2003cod}. On the other hand b) has been investigated in the framework of coevolving networks \cite{gross2008acn, perc2010cgm, zimmermann2004cds, ehrhardt2006pms, vazquez2007tsc, pacheco2008rgd, poncela2008ccn}, where the connections between individuals are not there forever but can be changed in parallel with the evolution of the opinions in order to increase the level of satisfaction in the system \cite{gil2006caa, suchecki2005vmd, holme2006npt, benczik2008lcs, nardini2008stf, vazquez2008gat, iniguez2009ocf}. 

Recently we have introduced a coevolving network model \cite{iniguez2009ocf, iniguez2010mof}, where not only short range ferromagnetic interactions  but also long range interactions were taken into account. This corresponds to the fact that, although our opinion is strongly influenced by our close friends, we are not independent of the general mood of the society. However, the impact of the society as a whole does not have to be ferromagnetic. As known from sociology again \cite{baron2010sps}, all individuals have two kinds of driving forces with respect to the society: We want to be similar to the average around us to use the society's protecting power and, at the same time, we want to be different to be distinguished as individuals. For every individual these conflicting components are present in different proportions, resulting in either net positive or net negative attitude with respect to the overall opinion of other individuals. This effect was taken into account \cite{iniguez2009ocf} by an attitude parameter $\alpha$ considered fixed or quenched to each individual. Since the attitude parameter can have positive or negative sign, it constitutes a source of frustration \cite{mezard1988sgt} in the system.

In \cite{iniguez2009ocf, iniguez2010mof} we also introduced a separation of time scales for different opinion formation mechanisms. While communications go on all the time leading to a quasi-continuous adjustment of the individuals' opinions, it takes more effort to make new friends than to quit with old ones. Therefore, we introduced a measure of time separation, which characterizes this difference by allowing for changes in the network neighborhoods after $g$ time steps of the difference equation governing the opinion update. We found interesting effects as a function of $g$ and the attitude parameter $\alpha$: For small values of $g$, where the rewiring process is very rapid and only two communities eventually develop, the attitude parameter plays a minor role and the $\alpha$ distributions in them were found to be broad and similar. However, for the intermediate values of $g$ the smaller communities have a rather narrow distribution with mostly negative $\alpha$ values, while the distributions for larger communities are broad and shifted toward positive $\alpha$ values.  Naturally, the agents with negative $\alpha$'s do not feel comfortable in a large homogeneous community, thus they tend to build smaller ones.

The aim of the present paper is to understand better the role of the attitude parameter and the separation of time scales in the coevolution of opinion and network structure of the underlying system. In order to do so, we define a model on a ring, and keep this topology preserved. Therefore, instead of rewiring the network we allow for location exchanges between agents by carrying their individual opinions and attitudes. This corresponds to a situation where the agent looks for a better environment to live in, and is reminiscent of Schelling's checkerboard model for residential segregation, where the relocation of agents with a mild preference for having a few alike neighbors in a static lattice can lead to fully segregated outcomes \cite{schelling1971dms, schelling1978mam, zhang2004rsa, vinkovic2006pas, stauffer2007iss, dall2008sps, gauvin2009pds}. The decision whether such an exchange is made is assumed dependent only on the short range interactions. However, in the opinion formation the attitude toward the social mood plays also a role. As the possibilities for finding new environments are limited, an amount of frustration will remain in the system for not too large values of the time separation parameter $g$. Interestingly, we see as a function of $g$ a rather sharp, phase transition-like change to a state without frustration as the individuals get enough time to form a firm opinion. 

This paper is organized as follows. In Section \ref{model} we introduce the model in detail. In Section~\ref{results} we present the numerical results. In Section~\ref{mean} a mean field calculation is presented, giving account for the variation in the number of undecided agents. Finally we draw conclusions.

\section{Model}
\label{model}
 
As in \cite{iniguez2009ocf, iniguez2010mof}, we study the dynamics of opinion formation in a network with a fixed number of individuals or agents ($N$) to whom a simple question is posed. For the network connectivity between agents, we here assume a 1d ring topology instead of a more complex network topology we studied earlier. A state variable $x_i \in [-x_{lim},x_{lim}]$ (for fixed $x_{lim} > 0$) is associated with each individual $i$, which measures the agent's instantaneous inclination concerning the question at hand, while the network links represent the presence of discussions between agents related to this question. 
The time scale for discussions or exchange of information between individuals (``transactions'') is $dt$, while the time scale for a generalized change of connections in the network (``generation'') is $T$. These two quantities are related by $T = gdt$, where the parameter $g$ defines the number of transactions per generation.
\medskip

The dynamics of the agent's state variable $x_i$ can be written as
\begin{equation}
\label{eq:microdyn}
\frac{\partial{x_i}}{\partial{t}} = f_s(\lbrace{x_j}\rbrace_s) x_i + f_l(\lbrace{x_j}\rbrace_l) \alpha_i,
\end{equation}
where the random parameter $\alpha_i \in [-\alpha_{lim},\alpha_{lim}]$ (for fixed $\alpha_{lim} > 0$) accounts for the agent's own attitude towards overall or public opinion. The short range interaction term $f_s(\lbrace{x_j}\rbrace_s) x_i$ describes the direct influence over $i$ of the subset of `close' agents $\lbrace{x_j}\rbrace_s$, while the long range interaction term $f_l(\lbrace{x_j}\rbrace_l) \alpha_i$ measures the indirect effect of the subset of `far' agents $\lbrace{x_j}\rbrace_l$ modulated by the attitude of $i$.
The system consist of a ring (a chain with periodic boundary conditions) where the short range interactions take place over the first $m$ neighbors of each agent, so the number of short range connections is $2m$. The long range interaction takes into account the average of opinion over the rest of agents in the network, that is,
\begin{align}
\label{eq:short}
f_s(\lbrace{x_j}\rbrace_s) x_i &= \left\langle x \right\rangle_{i}^{(m)} \text{sgn}(x_i) x_i = \frac{|x_i|}{2m} \sum_{\ell = 1}^{m} x_{i \pm \ell}, \\
\label{eq:long}
f_l(\lbrace{x_j}\rbrace_l) \alpha_i &= \left\langle x \right\rangle_{i}^{(N-m)} \alpha_i = \frac{\alpha_i}{N-1-2m} \sum_{\ell = m+1}^{[N/2]} x_{i \pm \ell},
\end{align}
where $\text{sgn}(x_i)$ denotes the sign of $x_i$. Observe that $m < (N - 1)/2$. Once the opinion of an agent reaches any of the limit values $\pm x_{lim}$, it stays fixed for the rest of the dynamics and the agent is said to be \textit{decided}. This is because we attempt to describe a state of total conviction that is unlikely to change anymore, like in a balloting process.

The dynamical evolution of the systems obeys Eq.~\ref{eq:microdyn} for $g$ time steps, when the agents are allowed to exchange places in the ring in order to help them reaching a definite decision ($|x_i| = x_{lim}$). This is done according to the following rules: One chooses $N^2$ pairs of agents at random, and  picks up the pairs of agents with both of them being not decided ($|x_i|, |x_j| < x_{lim}$). For these pairs one calculates a measure of the distance between the agents' opinions

\begin{equation}
\label{eq:rewBefore}
p_{ij}^{(m)} = \frac{1}{4 x_{lim}} \left[ |x_i - \left\langle x \right\rangle_{i}^{(m)}| + |x_j - \left\langle x \right\rangle_{j}^{(m)}| \right],
\end{equation}
and compares it with the same quantity if one exchanges $i$ and $j$, namely

\begin{equation}
\label{eq:rewAfter}
q_{ij}^{(m)} = \frac{1}{4 x_{lim}} \left[ |x_i - \left\langle x \right\rangle_{j}^{(m)}| + |x_j - \left\langle x \right\rangle_{i}^{(m)}| \right].
\end{equation}

If $p_{ij}^{(m)}>q_{ij}^{(m)}$ one exchanges places. In the above formulas {$\left\langle ... \right\rangle_{i}^{(m)}$ means the average over the $m$ left and right neighbors of site $i$, as in Eq.~\ref{eq:short}. This procedure is repeated every $g$ time steps, until one is not able to find favorable changes or all the agents have reached either one of the two limit opinions, since the exchange process only deals with pairs of undecided agents. Observe that these rules tend to increase opinion homogeneity in the system, which is reminiscent of the homophily principle mentioned in the previous section. A descriptive diagram of the system and the exchange process is shown in Fig.~\ref{fig:diagram}.

\begin{figure}[ht!]
\centering
\includegraphics[width=0.95\columnwidth]{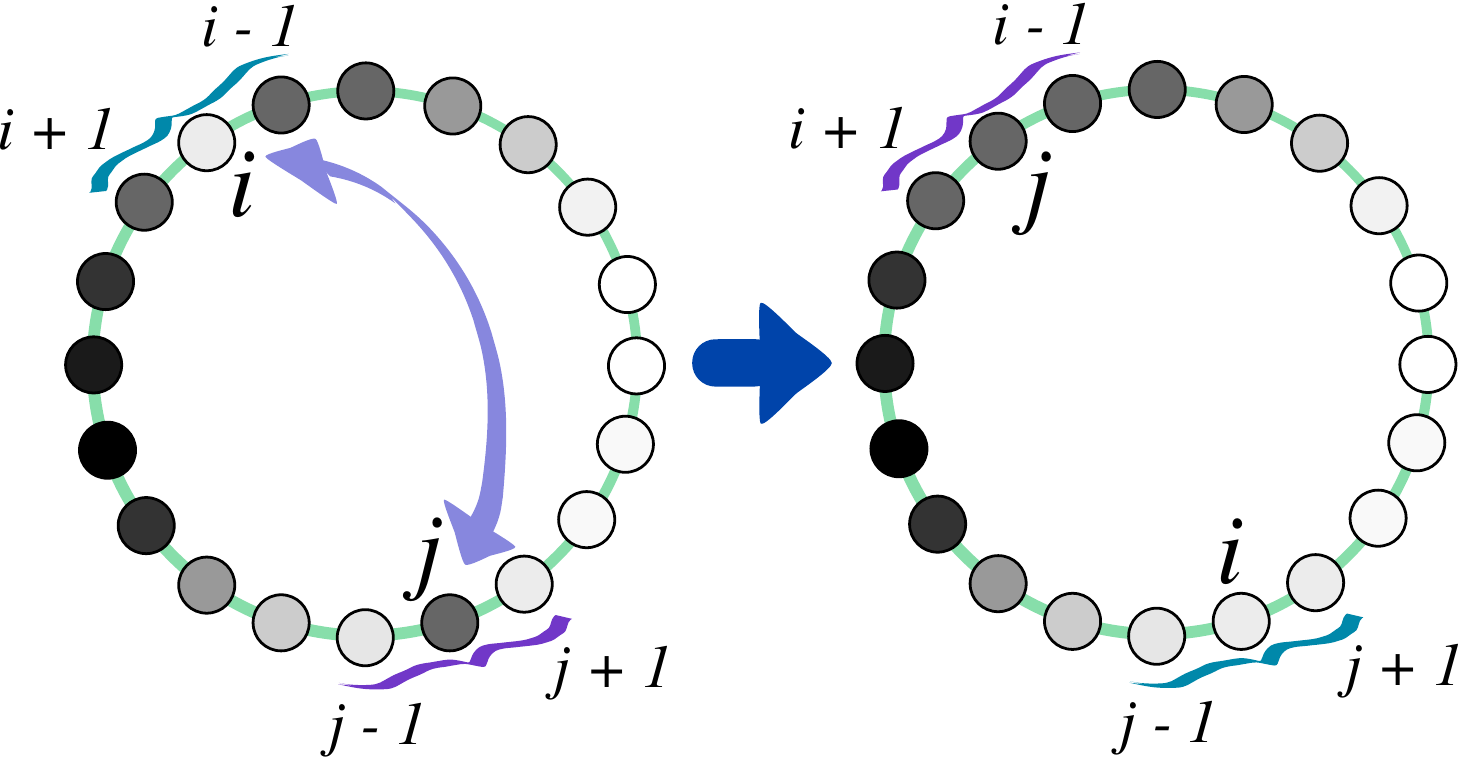}
\caption{(Color online) Diagram showing the exchange process used in the model (for $m=1$). The randomly chosen undecided nodes $i$ and $j$ have $q^{(1)}_{ij} < p^{(1)}_{ij}$, thus they are exchanged. Observe that the grey scale representing the opinion variable is more uniform after exchange.}
\label{fig:diagram}
\end{figure}

\section{Numerical Results}
\label{results}

We solve the model system by numerical simulations. For that the system is initialized with values of the state variable $x_i(0)$ in the interval $[-x_{lim}, x_{lim}]$ drawn randomly from the Gaussian distribution with zero mean and unit standard deviation, cut off at $\pm x_{lim}$ and $x_{lim} = 1$. Likewise the attitude parameter $\alpha_i$ for each agent was chosen randomly from a uniform distribution in the interval $[-\alpha_{lim},\alpha_{lim}]$ with $\alpha_{lim} = 1$ and kept fixed throughout the whole simulations.

The simulations have been carried out using a two-step process: first we solve the dynamical equations of opinion for all undecided agents by using a simple Euler numerical integration with time step $dt = 10^{-4}$, and then we perform the exchange process every $g$ time steps according to the above described rules. We keep track of the progress of the dynamics with two counters, namely the number of undecided agents ($n_{und}$) and the number of pairs that have been exchanged ($n_{exch}$) after every exchange process. As the agents reach the definite opinions, the counter $n_{und}$ will decrease from its initial value $N$ to some number close to zero. The asymptotic stationary value of $n_{und}$ is considered as the final number of undecided agents. Since the exchanged pairs have to be undecided, the counter $n_{exch}$ usually stays around or below $n_{und}^2$. The exchange process is realized sequentially and randomly such that agents can be chosen more than once in the same generation. However, the probability of such event decays fast with $N$ to be very rear to have an effect on the results.

In the simulations the dynamics is let to run until the exchange of agent locations takes place very rarely. The relaxation time for this is exceedingly large, and comparing the results of calculations with a large number of iterations we found that after $10^7$ transactions, the averaged results over 100 realizations differ by less that 0.1\%. Therefore, in all the calculations presented here we have used these numbers. Moreover, since it turned out that some results depend strongly on the size of the ring for small values of $N$, we chose to do most of the calculations on a ring of $N=5000$ and for the case $m=1$, i.e. the short range being limited to nearest neighbors. Finally, the exchange process rules can be relaxed so that any agent (decided or undecided) can be moved, yet since the exchange of decided agents is so seldom as to have any noticeable effect on the averaged results, we chose not to do so in order to increase the speed of the algorithm, especially for late stages of the dynamics.

\begin{figure}[h!]
\centering
\includegraphics[width=0.95\columnwidth]{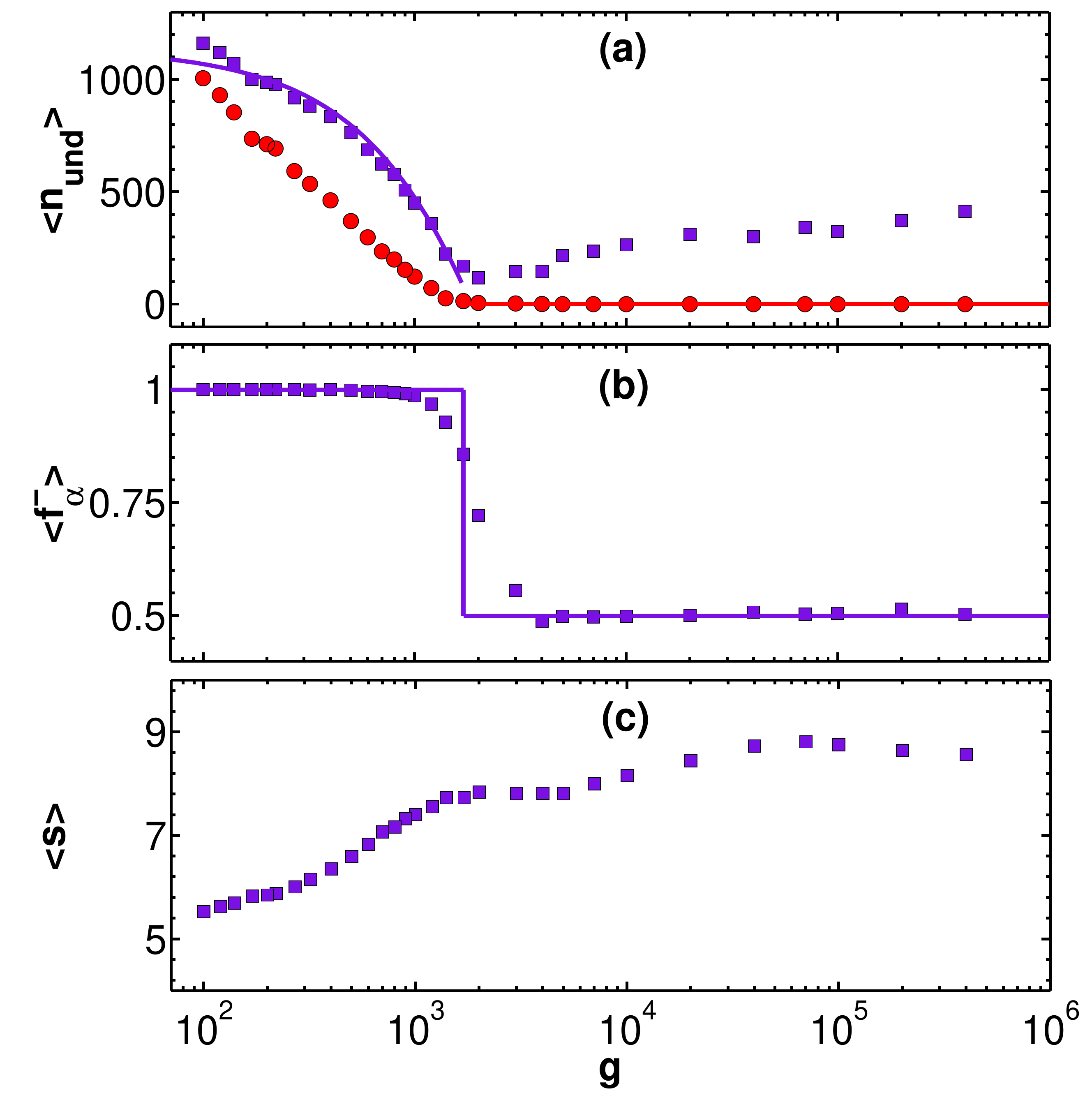}\
\caption{(Color online) (a) Average number of undecided agents as a function of the number of transactions per generation $g$. The purple squares are the numerical results after $10^7$ transactions and the red circles are the agents who will never get a decision, according to a linear analysis. The corresponding mean field predictions are shown as continuous lines. (b) The average fraction of undecided agents with negative $\alpha$ (purple squares) and its corresponding mean field prediction (continuous line).  (c) Average cluster size as a function of $g$. Observe the plateau around $g_c \approx 1.7 \times 10^3$.}
\label{fig:und}
\end{figure}

In Fig.~\ref{fig:und}(a) we show the average number of undecided agents ($\langle n_{und}\rangle$) in the ring as a function of the parameter $g$. It is clearly seen that there is a quite sharp minimum at $g_c \approx 1.7 \times 10^3$, a critical value of $g$ that can be predicted by using the mean field analysis, see Section~\ref{mean}. In the figure we also show by red circles the expected value of $\langle n_{und}\rangle$ for $t \to \infty$ as obtained from a linear analysis, and by continuous lines the corresponding mean field predictions, explained below in more detail. The critical value $g_c$ signals a change of phase in the system: for $g > g_c$ \textit{all} the agents get decided in the limit of infinite time, while for $g < g_c$ a finite fraction of the network remains undecided for arbitrarily long times. We identify the former phase as a state of maximum relaxation, and the latter as a frustrated state where many agents cannot reach the limit opinions.

In Fig.~\ref{fig:und}(b) it is seen that the average value of the fraction of undecided agents with {\it negative} $\alpha$ ($\langle f^-_{\alpha} \rangle$) also shows a sharp change of behavior, predicted by mean field as a continuous line. For $g > g_c$ the value is 1/2, that is, the undecided agents have positive and negative $\alpha$'s indistinctly, but for sufficiently small $g$-values most undecided agents have $\alpha < 0$. Such phase change behavior is also structural as evidenced in Fig.~\ref{fig:und}(c) where the average cluster size ($\langle s \rangle$) remains constant around the critical value $g_c$. Here we have defined a cluster as a set of connected agents having opinions of the same sign, independent whether they are decided or not. The maximum observed in Fig.~\ref{fig:und}(c) and likely the minimum in Fig.~\ref{fig:und}(a) are due to relaxation problems. This can be understood due to the dynamics being stopped at a fixed time for all $g$, which is not enough to reach the asymptotic state. 

\begin{figure}[h!]
\centering
\includegraphics[width=\columnwidth]{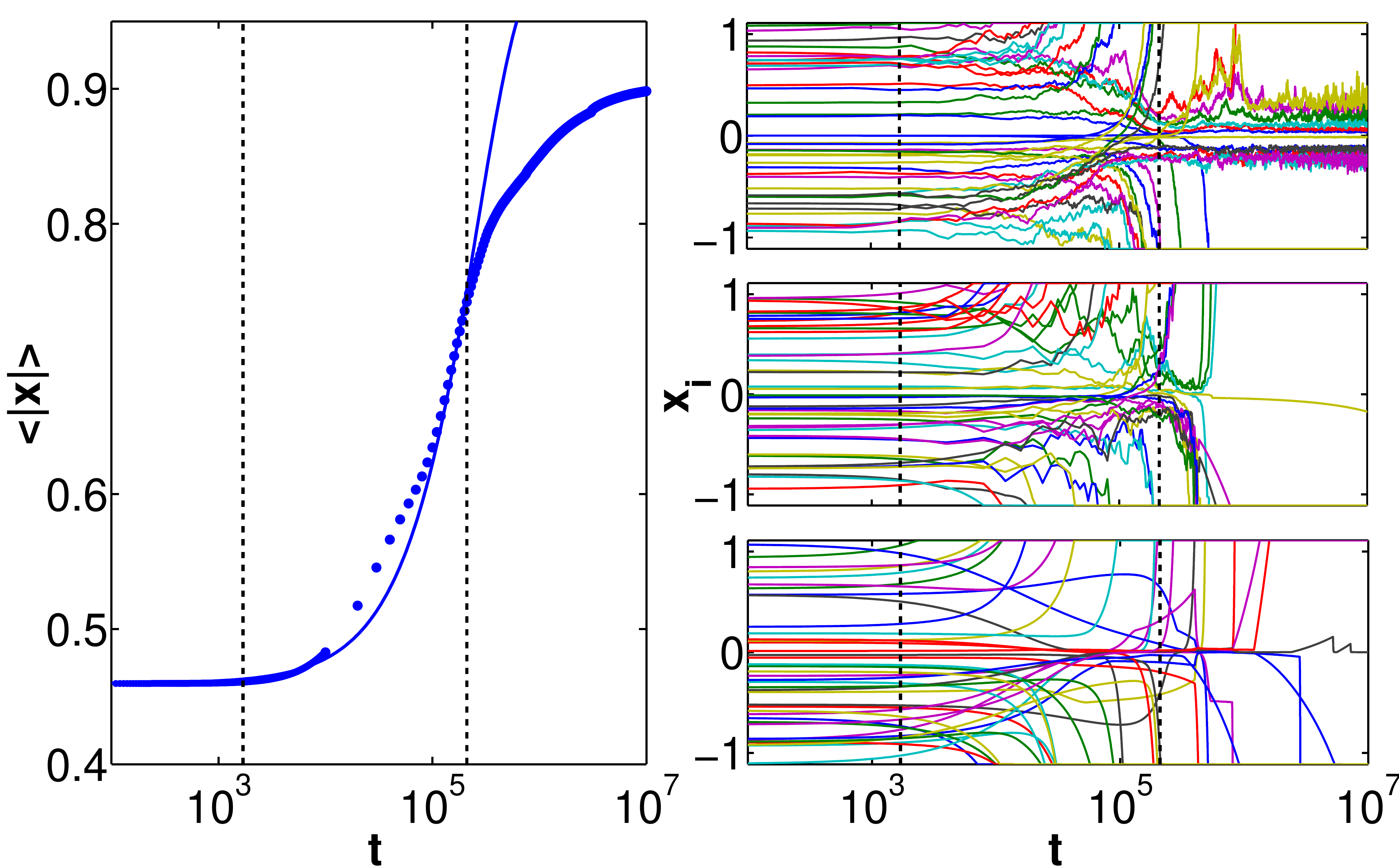}
\caption{(Color online) Left panel: Time evolution of the absolute value of opinion averaged over 100 realizations of the ring of $N = 5000$ agents for $g \to \infty$. The corresponding mean field prediction is shown as a continuous line. Right panels: Time history for a sample of 40 agents in a single realization for three different values of $g = 7 \times 10^2, 4 \times 10^3, 4\times 10^5$ (ordered from top to bottom), chosen to correspond to $g < g_c$, $g \sim  g_c$, and $g > g_c$, around $g_c$ obtained from Fig.~\ref{fig:und}. The vertical dotted lines indicate the moments when there is a change of time regime in the dynamics, as explained in the text.}
\label{fig:xm}
\end{figure}

On the left hand side of Fig.~\ref{fig:xm} we present the results for the relaxation dynamics, by plotting the average absolute value of the state variable $x$ as a function of time, when the exchange process is off ($g \to \infty$). The results are averages  over 100 realizations. It is clearly noticeable that there are three different time regimes, seen as an s-shape curve and predicted by a mean field treatment (see Section~\ref{mean}). Up to around $t \approx g_c$ (in units of $dt$) the evolution of $\langle |x| \rangle$ is very slow; between this value and $\tau \approx 2.1 \times 10^5$ the curve is concave upwards; finally for long times the variable approaches the asymptotic value $x_{lim} = 1$ very slowly and the curve is concave downwards. Only the evolution up to $10^7$ time steps is shown, where $\langle |x| \rangle \approx 0.9$ is reached.

On the right hand side of Fig.~\ref{fig:xm} we show the time history of a sample of 40 agents for a given realization and different values of $g$. If $g < g_c$ (top plot) the exchanges happen in the initial time regime, up to $t \approx g_c$. Therefore the evolution of the system cannot reach the relaxed state, and frustration appears in the form of indefinitely undecided agents. On the other hand, if $g\gtrsim g_c$ (middle and bottom plots) the relaxation is already advanced when the exchanges are carried out, which contribute to further relaxation. The fact that there is a minimum in the number of undecided nodes is due to the slow relaxation for larger $g$ values.

The separation of the three time regimes mentioned above is even clearer in these plots: 1) in the first time regime practically all the agents change their opinions slowly regardless of the value of $g$, so $t \approx g_c$ can be recognized as the characteristic time for the first agents in the network to get decided; 2) in the second time regime the dynamics speeds up exponentially and most agents get decided (with individual trajectories getting smoother as $g$ increases due to less frequent exchanges); 3) in the third and final time regime only some agents remain undecided, which for $g < g_c$ occupy frustrated regions in the network and will be undecided indefinitely, and for $g > g_c$ will get decided after a large but finite amount of time. These remarks are supported by the analytical treatment in Section~\ref{mean}.

\begin{figure}[h!]
\centering
\includegraphics[width=\columnwidth]{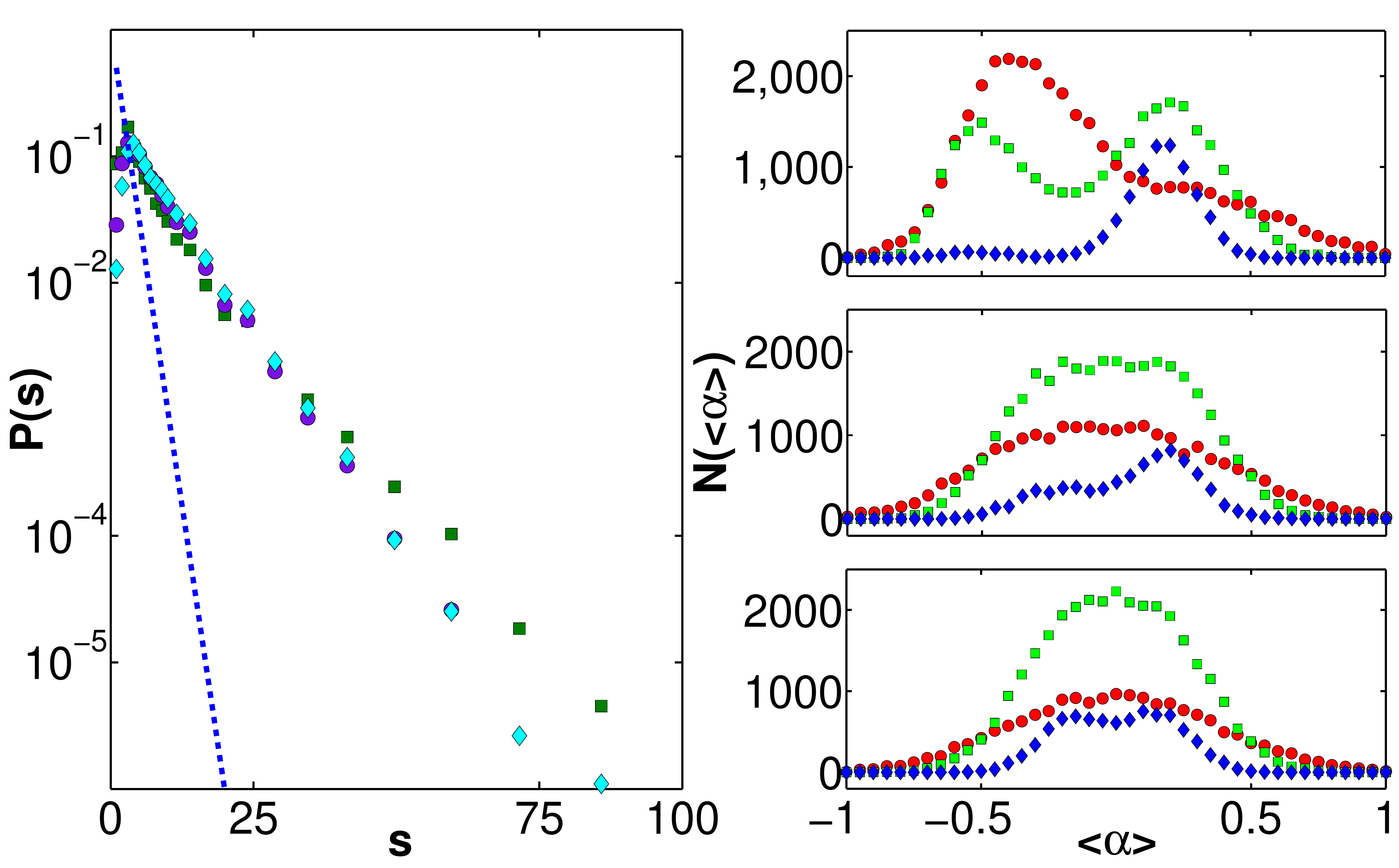}
\caption{(Color online) In the left hand side we show a semi-log plot of the cluster size distribution for three values of $g = 7 \times 10^2, 4 \times 10^3, 4\times 10^5$ (green squares, purple circles and light blue diamonds respectively). The distribution for the initial random ring is shown for comparison as a dotted blue line. The panels on the right hand side show the number distribution of $\alpha$ for 100 realizations of the same three values of $g$ (ordered from top to bottom) and for three different ranges of cluster size $s$: red circles for $s \in [1, 5)$, green squares for $s \in [5, 15)$, and blue diamonds for $s \in [15, 5000]$.}
\label{fig:sa}
\end{figure}

The cluster size distribution in the initial random ring goes as $P(s) = 1/2^s$ (e.g. the probability of having $s$ consecutive agents with the same sign of their initial opinions) and it is shown as a dotted blue line in the left panel of Fig.~\ref{fig:sa}. The cluster size distribution after $10^7$ transactions (also shown) turns out to change very little with $g$, although it is quite different from the random value. The phase change behavior seen by using $\langle n_{und} \rangle$ as an ``order parameter'' is also reflected in the preferred value of $\alpha$ for clusters of different size. In the right panels of Fig.~\ref{fig:sa} we show this effect by plotting the number distribution of $\alpha$ ($N(\langle \alpha \rangle)$) for the three values of $g$, i.e.  $g < g_c$, $g \sim  g_c$, and $g > g_c$, and for three different ranges of cluster size. Observe that for $g < g_c$ small clusters (of size 1 to 4) are composed mainly of agents with negative $\alpha$, clusters of medium size (5 to 14) present a bimodal distribution of negative and positive values, and large clusters (15 to 5000) have agents with positive $\alpha$. For $g$ around the critical value the picture changes dramatically and large clusters start having agents with $\alpha < 0$. For $g > g_c$ the number distribution of $\alpha$ approaches a Gaussian form independently of the cluster size.

So far we have considered the case where the short range interaction deals with nearest neighbors only ($m = 1$). We have also studied the situation in which the short range interaction includes the second neighbors, i.e. $m = 2$. In this case a phase change behavior is also clearly visible in the number of undecided agents as a function of $g$, though the position where it appears has moved slightly.

\section{Mean Field Calculations}
\label{mean}

In this section we shall investigate the peculiar features of the phase change exhibited by our model, namely, the reasons why some agents are undecided, the factors that determine the average cluster size, and the peculiar distribution of agents with negative attitude parameter $\alpha$ in the final network configuration. For this we shall perform a linear analysis of the dynamics and introduce some mean field ideas that may help understanding the role of the different time scales and their effect on the structure of the network, in particular the role of parameter $g$.

\subsection {Linear analysis}
\label{linstab}

The quantity that here plays the role of ``order parameter'' is the number of undecided agents. However, in any long but finite numerical calculation, out of the total number of agents that appear as undecided (see the purple squares in Fig.~\ref{fig:und}(a)) only a fraction will remain undecided forever. We shall investigate first the circumstances that prevent agents to reach a limit opinion.

There are different scenarios depending on the values of $f_s$ and $f_l$ in Eq.~\ref{eq:microdyn}. When the long range term $f_l = 0$, Eq.~\ref{eq:microdyn} has a simple exponential solution and the only situation that prevents the limit value $\text{sgn}(x_i)$ to be reached is when the short range term $f_s \leq 0$, so the agent remains undecided forever. If $f_s = 0$ and $f_l \neq 0$ the solution is linear in time and the agent will eventually reach a limit value. Notice that for $m = 1$ (i.e. short range interaction with nearest neighbors only) this situation corresponds to an agent at the border between two groups of opposite opinion, and once that agent becomes decided the border is displaced by one site. Since the ring is symmetric, the net displacement of the border will be zero, and this will give a characteristic cluster size. 

Eq.~\ref{eq:microdyn} exhibits various fixed points, on top of the limit values $x_i = \pm 1$. For each agent $i$ there is a fixed point at $|x_{0, i}| < 1$, where
\begin{equation}
\label{eq:fp}
x_{0, i} = -\frac{ f_l \alpha_i }{ f_s }.
\end{equation} 
If both $f_s$ and $f_l \neq 0$, one can perform a linear stability analysis around the fixed point of Eq.~\ref{eq:fp}. Then, agent $i$ is considered indefinitely undecided if this fixed point is stable, that is, when the real part of the eigenvalue
\begin {equation}
\label{eq:linear}
\lambda_i = \left. \frac{\partial (\partial_t x_i)}{\partial x_i}\right |_{x_i = x_{0, i}} = \langle x \rangle_{i}^{(1)}[2\theta(x_{0, i}) - 1],
\end{equation}
is negative. In this equation $\theta(x_{0, i})$ is the Heaviside step function. It should be noted that the occurrence of $\Re [\lambda_i] < 0$ is extremely rare without an exchange process (e.g. for $g \to \infty $). The reason is that $\text{sgn}(x_{0, i})$ must be opposite to the sign of $\langle x \rangle_{i}^{(1)}$, which has to be different from zero and eventually $\pm 1$. This means that the agent is embedded in a very adverse environment of immediate neighbors, a situation not favored by the dynamics that tends to diminish disagreement between the agent and its first neighbors. The only possibility for an agent to remain undecided forever is when the magnitude of $\alpha_i$ is large enough to hamper the dynamics. However, the $\alpha$ distribution is flat and the probability for this to happen is of the order of $\mathcal{O}(1/N)$.

Summarizing, an agent can only be undecided in the limit of $t \to \infty$ if:
\begin{enumerate}
 \renewcommand{\labelenumi}{(\alph{enumi})}
 \item $f_l = 0$ and $f_s \leq 0$, or if 
 \item $f_s, f_l \neq 0$ and $\Re [\lambda_i] < 0$. 
\end{enumerate}
From the value of $\langle n_{und} \rangle$ after $10^7$ transactions (shown in Fig.~\ref{fig:und}(a) as purple squares), we have tested all agents that fullfil any of these two conditions to remain undecided forever, and plotted their numbers in the figure as red circles. Indeed, the asymptotic number of undecided agents is nonzero for $g < g_c$ and zero for $g > g_c$. The latter is in agreement with our previous mean field prediction, drawn as a continuous red line in Fig.~\ref{fig:und}(a).
 
We now investigate the form of the curve for the number of undecided agents for $g < g_c$, which can be estimated from the initial Gaussian distribution of $x$. First, the symmetry of sign in the distribution of $\alpha$ implies that only half of the agents are likely to have $\alpha < 0$ and thus be undecided. Then, since Eq.~\ref{eq:microdyn} has an approximate solution $x = x(0) e^{t/g_c}$ before the exchange process takes place, at least those agents with initial $|x(0)| < x_g$ remain undecided at $t = g$, where $x_g = e^{-g/g_c}$. Therefore, the number of undecided agents as a function of $g$ can be calculated from the initial distribution of $x$ as
\begin{equation}
\label{eq:meanUnd}
n_{und}(g) = \frac{N}{2} \frac{\mathrm{erf}(e^{-g / (g_c \sqrt 2)} / \sqrt 2) - \mathrm{erf}(e^{-1 / \sqrt 2} / \sqrt 2)}{\mathrm{erf}(1/\sqrt 2)}
\end{equation}
where $\mathrm{erf}(x) = (2/\sqrt{\pi}) \int_0^x e^{-u^2} du$ is the error function, and the factor of 1/2 is due to the sign symmetry. The result of Eq.~\ref{eq:meanUnd} is plotted in Fig.~\ref{fig:und}(a) as a purple line, where the value $g_c \approx 1.7 \times 10^3$ has been fitted with least-squares technique. Notice that the agreement with the calculation (purple squares) is considerably good. The theoretical estimation of the truly undecided agents (red circles in the figure) is more involved, since the actions of exchanging become important, and this will be the matter of further study.
 
\subsection{Mean field for \texorpdfstring{$\langle |x| \rangle$}{<|x|>}}
\label{meanOpi}

The time evolution of the average absolute value of opinion in the network when there are no exchanges can be understood by an estimation of the characteristic time ($\tau$) for the whole system to reach the limit values of opinion. This is done assuming that $g \gtrsim \tau$, where one can by use a mean field approach similar to the one described in our previous model \cite{iniguez2009ocf}. Although the network topology there is different, the mechanisms that result in the magnetization relaxation of all ferromagnetic-like problems are similar. The average number of undecided agents as a function of time is found to be
\begin{equation}
\label{eq:brillouin}
\langle n_{und}(t) \rangle = N - (N + 1) \coth \left( \frac{N + 1}{2} \frac{t}{\tau} \right) + \coth \left( \frac{t}{2\tau} \right),
\end{equation}
where the time scale $\tau$ is related to the critical value $g_c$ as $\tau = g_c N/40$, see \cite{iniguez2009ocf}. We now follow a procedure similar to that of the previous subsection, without considering exchange processes. Since Eq.~\ref{eq:microdyn} has an approximate solution $x = x(0) e^{t/\tau}$, only the agents with initial opinion $|x(0)| > x_t = e^{-t/\tau}$ can get decided at time $t$, while the rest of the agents are still undecided. Then the average absolute opinion of the decided agents is $2 \int_{x_t}^1 P(x)dx$ and that of the undecided agents is $2 e^{t/\tau} \int_{0}^{x_t} xP(x)dx$. By integrating the distribution of initial opinions $P(x)$ we get
\begin{equation}
\label{eq:opinion}
\langle |x(t)| \rangle = 1 - \frac{\mathrm{erf}( x_t/\sqrt{2} )}{\mathrm{erf}( 1/\sqrt{2} )} + \frac{\sqrt{2/\pi}}{\mathrm{erf}( 1/\sqrt{2} )} e^{t/\tau} \left( 1 - e^{-x_t^2/2} \right).
\end{equation}

Eq.~\ref{eq:opinion} has been fitted to the numerical results shown in the left panel of Fig.~\ref{fig:xm} with least-squares technique, giving a value of $\tau \approx 2.1 \times 10^5$, which in turn corresponds to $g_c = 40 \tau / N \approx 1.7 \times 10^3$. This is in good agreement with our estimate of last subsection and with the value in Fig.~\ref{fig:und}(a) of $g$ at the minimum in the number of undecided agents after a finite number of transactions. Moreover, the slope of Eq.~\ref{eq:opinion} is $\langle |x| \rangle' \approx 2 \times 10^{-6}$ for $0 < t < g_c$, then it drops fast at around $t \approx \tau$ and is asymptotically zero for $t \gg \tau$. This illustrates the three time regimes of opinion evolution discussed in Section~\ref{results} and indicated in Fig.~\ref{fig:xm} as vertical dotted lines. We detected that the fitting of Eq.~\ref{eq:opinion} is very good for short times but starts to deviate significantly for longer times, in a similar fashion as the approximation for the fraction of undecided agents in \cite{iniguez2009ocf}. This is to be expected, since in this mean field approach we have not taken into account the effects produced by the random distribution of $\alpha$. As a consequence Eq.~\ref{eq:opinion} relaxes faster to the asymptotic state $\langle |x| \rangle = 1$ than the actual dynamics.

\subsection{Analysis of \texorpdfstring{$\langle f^-_{\alpha} \rangle$}{<f>} and \texorpdfstring{$\langle s \rangle$}{<s>}}
\label{meanAlphSize}

From the subset of undecided agents after a long but finite time we can also calculate the fraction of agents with negative $\alpha$. For $g > g_c$, the symmetry of sign in Eq.~\ref{eq:microdyn} and in the initial $x$ and $\alpha$ distributions implies that $\alpha$ should be distributed evenly among all agents and a value of $\langle f^-_{\alpha} \rangle = 1/2$ is predicted. This agrees very well with the numerical results, and is depicted as a continuous purple line in the right part of Fig.~\ref{fig:und}(b). 

On the other hand, for $g < g_c$ all undecided agents have $\alpha < 0$. This fact, although apparently logical, is puzzling, since it holds even for reasonably large values of $g$ (up to $10^3$), but it can be explained as follows: After running the dynamics for a long time $f_l$ is a very small number, since the average overall opinion approaches zero, and the second term of Eq.~\ref{eq:microdyn} is no longer important. Therefore, the only way that agent $i$ avoids the exponential approach to a limit opinion and remains undecided is that it finds itself in an adverse environment, such that it is likely to be chosen for an exchange many times. Furthermore, the exchanges have to modify the tendency of the agent towards a given limit opinion constantly. Remember that the condition to be chosen for exchange is $p_{ij}^{(1)} > q_{ij}^{(1)}$, meaning that the opinion in the neighborhoods of agents $i$ and $j$ is more homogeneous after the exchange.

We now show that only the agents with negative $\alpha$ can be in this situation after a large number of exchanges have taken place. Consider an exchange process between two undecided agents with the same sign in their attitude parameter. If $\alpha_i$ is positive the dynamics of Eq.~\ref{eq:microdyn} makes it likely that agent $i$ is surrounded by neighbors that share its own opinion, and since agent $j$ is in the same conditions as agent $i$, one infers from Eq.~\ref{eq:rewBefore} that $p_{ij}^{(1)} \approx 0$. Since there is always a possibility that the opinion in the neighborhoods of the two agents have opposite sign, from Eq.~~\ref{eq:rewAfter} one gets $q_{ij}^{(1)} > 0$ and thus agents $i$ and $j$ are very seldom chosen for a location exchange, eventually reaching limit values of opinion. On the other hand, if $\alpha_i$ is negative it is likely that $x_i$ and $\langle x \rangle_i^{(1)}$ have opposite signs, and a similar effect in the neighborhood of agent $j$ results in $p_{ij}^{(1)} > 0$. Since approximately half of the agents $j$ share the same sign as the neighbors of $i$ and viceversa, $q_{ij}^{(1)} \approx 0$ and the exchange can be performed. It should be noted that a similar analysis holds when the agents have opposite signs in their attitude parameter, thus $p_{ij}^{(1)} \approx q_{ij}^{(1)}$ and the exchanges are not as common as when both $\alpha_i$ and $\alpha_j$ are negative. Therefore, a negative attitude parameter along with the existence of many exchange processes hampers the possibility of agents attaining a definite decision. This result is plotted in the left part of Fig.~\ref{fig:und}(b) as a horizontal purple line.

As a corollary of this analysis, we can anticipate the structure of a typical configuration of opinions in the system after a large but finite number of transactions. For $g < g_c$ the undecided agents with $\alpha < 0$ have been exchanged many times in a random fashion and thus form small groups between large clusters of decided agents, as confirmed by visual inspection of single realizations of the numerical calculations. For $g > g_c$ most undecided agents have $f_s = 0$ and evolve slowly and linearly towards a limit opinion, therefore they should be at the borders of clusters with different definite opinions. This is truly the case, as can be seen by comparing the purple squares on the right hand side of Fig.~\ref{fig:und}(a) with the corresponding purple squares of Fig.~\ref{fig:und}(c). Indeed, the average cluster size is $\langle s \rangle \approx 10$ for large $g$, and the number of undecided agents detected in the calculation is $\langle n_{und} \rangle \approx 500$, which is approximately $N/\langle s \rangle$.

\section{Discussion}
\label{disc}

In this paper we studied the coevolution of opinions and the embedding of individuals in their environment. For the opinion dynamics we adopted earlier introduced continuous state variable equations \cite{iniguez2009ocf}, that include short range ferromagnetic interactions for describing homophily between neighboring agents, and long range interactions for describing how the overall mood of the majority affects the agent modulated by its attitude parameter being either positive or negative. This opinion update gives rise to short time scale transaction dynamics. For the model geometry or connectivity between agents, we used ring topology instead of a more complex network topology, we studied earlier \cite{iniguez2009ocf, iniguez2010mof}. The long or slow time scale dynamics of environment changes was carried out by exchanging the locations of pairs of agents. These two time scales are then well separated and their relation serves as a control parameter. 

As the main result of our study we find that by varying the time-scale parameter there is a phase change in the number of undecided individuals, which turned out to be mainly driven by the environment exchange dynamics. In order to understand this effect the following should be noted. First there is competing interaction due to the negative $\alpha$'s. Second, due to the asymmetry between the long and short range interactions (since only the latter are considered for exchanges), this competition does not lead to permanent frustration, provided that enough time is given for relaxation. However, if the relaxation is hampered by too frequent changes in the neighborhood as well as by not allowing enough exchanges to find the global optimum, frustration appears as a nonzero number of indefinitely undecided agents. Thus the phase change behavior is due to the separation of time scales and due to insufficient relaxation. The mean field analysis which we performed for the system supports the above picture.

It should be noted that there are relevant similarities and differences between the ring model studied here and our previous network model \cite{iniguez2009ocf}. First, the transaction dynamics defined by Eq.~\ref{eq:microdyn} is equivalent in both models, therefore producing similar relaxation processes in the limit of large $g$ that can be studied with the same mean field approach, as has been discussed in Section~\ref{mean}. Second, the exchange process used in the ring model is systematically different from the rewiring framework of the network model, since it changes the opinion distribution in the system by keeping the topology constant, thus making both models fundamentally distinct. Furthermore, the exchange process used here has some sociological background with the concept of homophily, since it favors homogeneity of opinion in the ring, but it is simple enough as to allow a deeper mean field treatment than the one performed for the network model.

Even though the exchange process differs from the rewiring scheme treated in \cite{iniguez2010mof}, both models can produce small clusters of agents with $\alpha < 0$ and large clusters of agents with $\alpha > 0$ for appropriate values of $g$, as shown in Fig.~\ref{fig:sa}. The mean field performed in the simpler ring topology suggests that a low time-scale ratio is responsible for frustration that creates small groups of undecided agents with negative $\alpha$, while a full relaxation tends to destroy this structure. In the case of the network model the rewiring rules would then be the sources of frustration, and an appropriate value of $g$ could freeze agents with $\alpha < 0$ in positions that create an heterogeneous community structure in the network. It is in this sense that a simpler though different ring model can give insight into the complex processes ocuring in our previous network model.

In this paper we have analyzed different aspects of opinion formation and have accordingly reached the view that the distinction between close relations and global influence is an important one. While for close friends homophily dominates, the attitude to the overall mood of the society can be both positive and negative. Due to the latter component competing interactions occur in the system. Another important aspect is that - similarly to physical phenomena - social interactions may take place at very different time scales. If the separation of time scales is strong, the system can relax and frustration becomes irrelevant, leading to a diminishing amount of undecided agents. When the two time scales approach each other the possibility of full relaxation vanishes, then frustration appears and a finite amount of undecided agents remain. We think that the phase change due to a variation in the separation of time scales is an interesting effect even beyond the present context.

\acknowledgments 

G.I. and K.K. acknowledge the Academy of Finland, the Finnish Center of Excellence program 2006 - 2011, under Project No. 129670. K.K. and J.K. acknowledge support from EU's FP7 FET Open STREP Project ICTeCollective No. 238597 and J.K. also support by Finland Distinguished Professor (FiDiPro) program of TEKES. K.K. and R.A.B. want to acknowledge financial support from Conacyt through Project No. 79641. R.A.B. is grateful to the Centre of Excellence in Computational Complex Systems Research - COSY of Aalto University for support and hospitality for the visits when most of this work has been done.

\bibliography{article}

\end{document}